\newif\ifdoubleblind
\newif\ifacm

\ifacm
	\documentclass[sigconf]{acmart}
\else
	\documentclass[conference]{IEEEtran}
\fi

\usepackage{amsmath}

\DeclareMathOperator{\diag}{diag}

\usepackage{amsmath,amsfonts,amssymb}
\usepackage{multirow}
\usepackage{tabularx,booktabs}
\usepackage{xspace}
\usepackage{balance}

\usepackage[binary-units=true]{siunitx}
\usepackage{tikz}
\usetikzlibrary{positioning}

\ifacm
	\usepackage[nolist]{acronym}
\usepackage{booktabs} 
\usepackage{subfig}
\usepackage{balance}

\usepackage{array}
\usepackage{graphicx}
\usepackage{wasysym}
\usepackage{xparse}

\usepackage{makecell}
\newcolumntype{Y}{>{\centering\arraybackslash}X}

\hypersetup{draft}

\renewcommand\footnotetextcopyrightpermission[1]{} 
\else
	\usepackage{graphicx}
\usepackage{epstopdf}
\usepackage{standalone}
\usepackage[nolist ]{acronym}
\usepackage{tcolorbox}
\usepackage{pdfpages}
\usepackage[bookmarks=false]{hyperref}
\hypersetup{draft}
\usepackage{pdfcomment}
\usepackage{hologo}

\usepackage{xstring}

\usepackage[utf8]{inputenc}
\usepackage{marvosym}

\usepackage{algorithm}
\usepackage{algpseudocode}

\usepackage{psfrag}
\usepackage{graphicx}

\usepackage{rotating}

\renewcommand{\baselinestretch}{1}

\usepackage[caption=false,font=footnotesize]{subfig}
\usepackage{stfloats}
\fi

\begin{document}

\newcommand{\paperTitle}{Machine Learning-Enabled Data Rate Prediction for 5G NSA Vehicle-to-Cloud Communications}
\newcommand{\paperAuthors}{Benjamin Sliwa, Hendrik Schippers, and Christian Wietfeld}
\newcommand{\paperEmails}{$\{$Benjamin.Sliwa, Hendrik.Schippers, Christian.Wietfeld$\}$@tu-dortmund.de}

\newcommand\single{1\textwidth}
\newcommand\double{.48\textwidth}
\newcommand\triple{.32\textwidth}
\newcommand\quarter{.24\textwidth}
\newcommand\singleC{1\columnwidth}
\newcommand\doubleC{.475\columnwidth}

\newcommand{\figurePadding}{0pt}
\newcommand{\figureTopPadding}{\figurePadding}
\newcommand{\figureBottomPadding}{\figurePadding}
\newcommand\red[1]{\colorbox{red}{\textbf{TODO:} #1}}

\newcommand\samsung{Samsung Galaxy S21 5G\xspace}
\newcommand\oneplus{OnePlus 8 Pro\xspace}
\newcommand\dortmund{\emph{Dortmund}\xspace}
\newcommand\hamm{\emph{Hamm}\xspace}
\newcommand\bonn{\emph{Bonn}\xspace}
\newcommand\cologne{\emph{Cologne}\xspace}
\newcommand\udp{\ac{UDP}\xspace}
\newcommand\tcp{\ac{TCP}\xspace}

\newcommand\tikzFig[2]
{
	\begin{tikzpicture}
		\node[draw,minimum height=#2,minimum width=\columnwidth,text width=\columnwidth,pos=0.5]{\LARGE #1};
	\end{tikzpicture}
}

\newcommand{\dummy}[3]
{
	\begin{figure}[b!]  
		\begin{tikzpicture}
		\node[draw,minimum height=6cm,minimum width=\columnwidth,text width=\columnwidth,pos=0.5]{\LARGE #1};
		\end{tikzpicture}
		\caption{#2}
		\label{#3}
	\end{figure}
}

\newcommand\pos{h!tb}

\newcommand{\basicFig}[7]
{
	\begin{figure}[#1]  	
		\vspace{#6}
		\centering		  
		\includegraphics[width=#7\columnwidth]{#2}
		\caption{#3}
		\label{#4}
		\vspace{#5}	
	\end{figure}
}
\newcommand{\fig}[4]{\basicFig{#1}{#2}{#3}{#4}{0cm}{0cm}{1}}

\newcommand\sFig[2]{\begin{subfigure}{#2}\includegraphics[width=\textwidth]{#1}\caption{}\end{subfigure}}
\newcommand\vs{\vspace{-0.3cm}}
\newcommand\vsF{\vspace{-0.4cm}}

\newcommand{\subfig}[3]
{%
	\subfloat[#3]%
	{%
		\includegraphics[width=#2\textwidth]{#1}%
	}%
	\hfill%
}

\newcommand\circled[1] 
{
	\tikz[baseline=(char.base)]
	{
		\node[shape=circle,draw,inner sep=1pt] (char) {#1};
	}\xspace
}
\begin{acronym}
	\acro{DDNS}{Data-Driven Network Simulation}
	\acro{LIMITS}{LIghweight Machine learning for IoT Systems}
	\acro{LIMoSim}{Lightweight ICT-centric Mobility Simulation}
	\acro{mmWave}{millimeter Wave}
	\acro{KNN}{k-Nearest Neighbors}
	\acro{RF}{Random Forest}
	\acro{NSA}{Non-Standalone}
	\acro{SA}{Standalone}
	\acro{WEKA}{Waikato Environment for Knowledge Analysis}
	\acro{LTE}{Long Term Evolution}
	\acro{MNO}{Mobile Network Operator}
	\acro{eMBB}{Enhanced Mobile Broadband}
	\acro{mMTC}{Massive Machine Type Communications}
	\acro{URLLC}{Ultra Reliable and Low Latency Communications}
	\acro{XGB}{Xtreme Grandient Boosting}
	\acro{SVM}{Support Vector Machine}
	\acro{ANN}{Artificial Neural Network}
	\acro{LR}{Linear Regression}
	\acro{UDP}{User Datagram Protocol}
	\acro{TCP}{Transmission Control Protocol}
	\acro{MAE}{Mean Absolute Error}
	\acro{RMSE}{Root Mean Square Error}
	\acro{5GAA}{5G Automotive Association}
	\acro{RSRP}{Reference Signal Received Power}
	\acro{RSRQ}{Reference Signal Received Quality}
	\acro{SINR}{Signal-to-Inteference-Plus-Noise Ratio}
	\acro{TA}{Timing Advance}
	\acro{RS}{Reference Signal}
	\acro{SS}{Synchronization Signal}
	\acro{QoS}{Quality of Service}
	\acro{RAT}{Radio Access Technology}
	\acro{RAN}{Radio Access Network}
	\acro{eNB}{evolved Node B}
	\acro{gNB}{next generation Node B}
	\acro{MDI}{Mean Decrease Impurity}
	\acro{API}{Aplication Programming Interface}
	\acro{SoC}{System-on-a-Chip}
	\acro{3GPP}{3rd Generation Partnership Project}
	\acro{UE}{User Equipment}
	\acro{COTS}{commertial off-the-shelf}
	\acro{CAGR}{Compound Annual Growth Rate}
	\acro{IoT}{Internet of Things}
	\acro{HD}{High Definition}
	\acro{ITS}{Intelligent Transportation System}
	\acro{RBF}{Radial Basis Function}
	\acro{ReLU}{Rectified Linear Unit}
\end{acronym}

\title{\paperTitle}

\ifacm
\else

	\title{\paperTitle}

	\ifdoubleblind
	\author{\IEEEauthorblockN{\textbf{Anonymous Authors}}
		\IEEEauthorblockA{Anonymous Institutions\\
			e-mail: Anonymous Emails}}
	\else
	\author{\IEEEauthorblockN{\textbf{\paperAuthors}}
		\IEEEauthorblockA{Communication Networks Institute,	TU Dortmund University, 44227 Dortmund, Germany\\
			e-mail: \paperEmails}}
	\fi
	
	\maketitle

\fi




\begin{abstract}

%
%
In order to satisfy the ever-growing \ac{QoS} requirements of innovative services, cellular communication networks are constantly evolving. Recently, the 5G \ac{NSA} mode has been deployed as an intermediate strategy to deliver high-speed connectivity to early adopters of 5G by incorporating \ac{LTE} network infrastructure.
%
%
In addition to the technological advancements, novel communication paradigms such as anticipatory mobile networking aim to achieve a more intelligent usage of the available network resources through exploitation of context knowledge. For this purpose, novel methods for proactive prediction of the end-to-end behavior are seen as key enablers.
%
%
In this paper, we present a first empirical analysis of client-based end-to-end data rate prediction for 5G \ac{NSA} vehicle-to-cloud communications. Although this operation mode is characterized by massive fluctuations of the observed data rate, the results show that conventional machine learning methods can utilize locally acquirable measurements for achieving comparably accurate estimations of the end-to-end behavior.

\end{abstract}

\begin{tikzpicture}[remember picture, overlay]
\node[below=5mm of current page.north, text width=20cm,font=\sffamily\footnotesize,align=center] {Accepted for presentation in: IEEE 4th 5G World Forum (5GWF) (WF-5G'21)\vspace{0.3cm}\\\pdfcomment[color=yellow,icon=Note]{
@InProceedings\{Sliwa2021machine,\\
	Author = \{Benjamin Sliwa and Hendrik Schippers and Christian Wietfeld\},\\
	Title = \{Machine Learning-Enabled Data Rate Prediction for \{5G\} \{NSA\} Vehicle-to-Cloud Communications\},\\
	Booktitle = \{IEEE 4th 5G World Forum (5GWF) (WF-5G'21)\},\\
	Year = \{2021\},\\
	Address = \{Virtual\},\\
	Month = \{Oct\},\\
\}
}};
\node[above=5mm of current page.south, text width=15cm,font=\sffamily\footnotesize] {2021~IEEE. Personal use of this material is permitted. Permission from IEEE must be obtained for all other uses, including reprinting/republishing this material for advertising or promotional purposes, collecting new collected works for resale or redistribution to servers or lists, or reuse of any copyrighted component of this work in other works.};
\end{tikzpicture}

\maketitle
\section{Introduction}

%
%
Data is anticipated to become the ``new oil'' of the automotive industry. While traditionally, moving vehicles were only regarded as means for personal transportation, their role within the \acp{ITS} is currently experiencing a significant transformation. Due to their various sensing and communication capabilities, cars are able to become moving sensor nodes and act as the main resources of novel data-driven services in the context of traffic sensing, \ac{HD} mapping of the geographic and the radio environment, as well as for distributed weather and air quality sensing.
%
%
As a consequence of the rise of the \emph{vehicular} big data paradigm, the resource optimization of the  vehicle-to-cloud communication has become an emerging research field \cite{Sliwa/etal/2021b, Sliwa/etal/2019d}.
%
%
According to a recent study by Cisco \cite{Cisco/2020a}, connected car services are expected to become the fastest growing type of \ac{IoT} connections in the next years with an expected \ac{CAGR} of \SI{30}{\percent}.

\begin{figure}[t]  	
	\vspace{0cm}
	\centering		  
	\includegraphics[width=1.0\columnwidth]{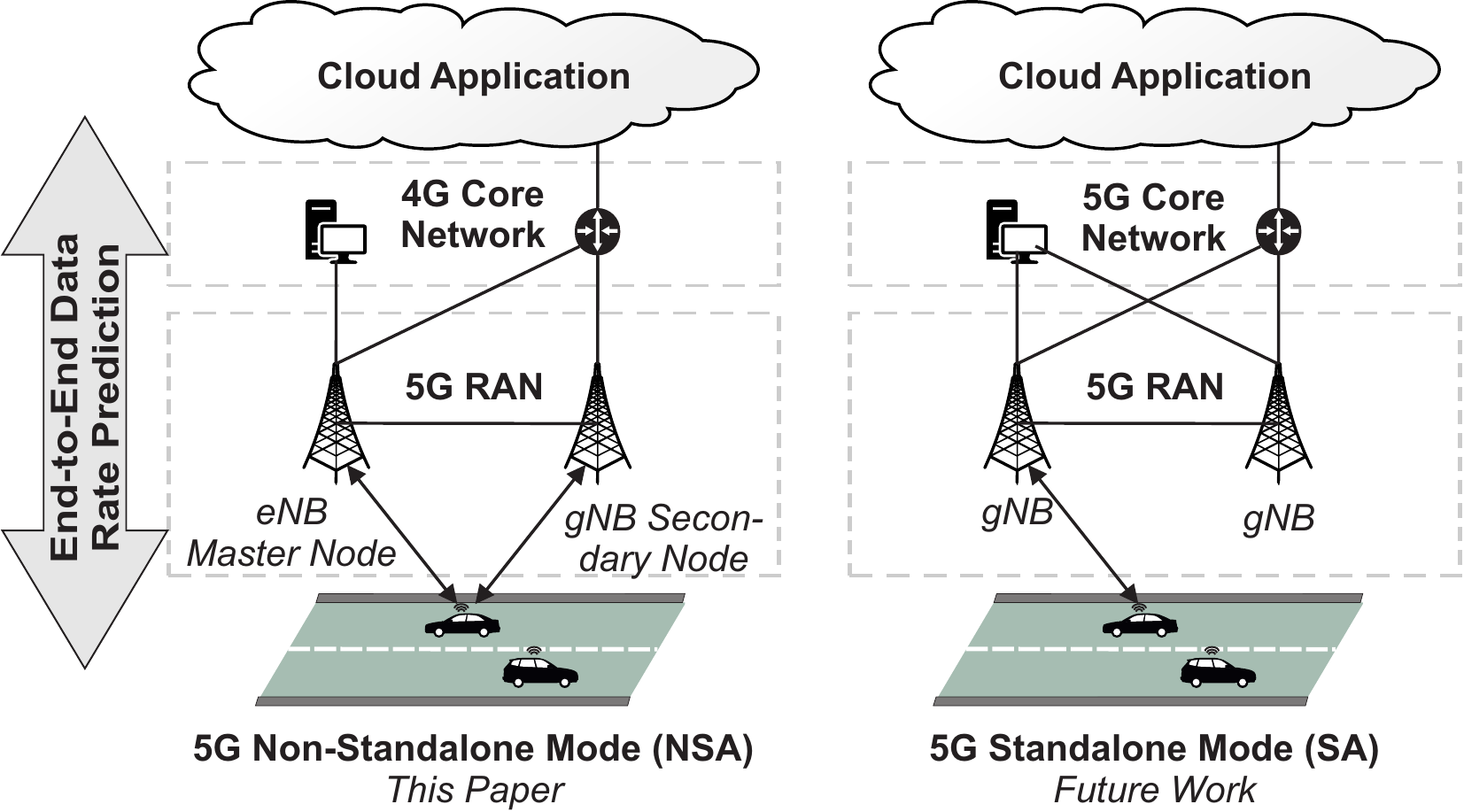}
	\caption{Schematic comparison of the 5G \ac{SA} and \ac{NSA} network architectures. This work focuses on 5G \ac{NSA} end-to-end data rate prediction from the mobile device through the public cellular network to an application server.}
	\label{fig:5g_nsa_sa}
	\vspace{0cm}	
\end{figure}

%
%
In parallel to the intensification of the resource demands and \ac{QoS} requirements of the novel applications and services, cellular communication networks are subject to a technology evolution that aims to achieve a more efficient usage of the shared radio channel and the limited spectrum resources. To this date, the deployment of the emerging 5G network technology has started worldwide.
%
%
While the 5G vision consists of the three major building blocks \ac{eMBB}, \ac{mMTC}, and \ac{URLLC}, most operators rely on a sequential approach for implementing these novel mechanisms. The 5G \ac{NSA} mode is an intermediate approach for offering the users \ac{eMBB}-compliant data rates. As shown illustrated in Fig.~\ref{fig:5g_nsa_sa}, this is realized through integrating new 5G \ac{RAN} components into existing \ac{LTE} networks.
%
%
In contrast to that, the full-featured 5G \ac{SA} mode will rely on a dedicated 5G core network that allows for future implementation of network slicing, \ac{mMTC}, and \ac{URLLC}.

%
%

%
%
Anticipatory mobile networking \cite{Bui/etal/2017a} is a novel communications paradigm that proposes the exploitation of context knowledge for proactive system and network optimization. As pointed out by a recent white paper of the \ac{5GAA} \cite{5GAA/2020a}, this approach is expected to become one of the key enablers for future connected and autonomous driving. Client knowledge about the end-to-end communication efficiency --- e.g., represented by the predicted end-to-end data rate --- allows the mobile clients to become parts of the network fabric and contribute to improving the overall network efficiency, e.g., via multi-\ac{RAT} networking \cite{Sepulcre/Gozalvez/2019a} and opportunistic data transfer \cite{Sliwa/etal/2019d, Sliwa/etal/2021b}.

%
%
In continuity to previous work \cite{Sliwa/Wietfeld/2019c}, which presented an empirical analysis of client-based data rate prediction in vehicular \ac{LTE} networks, this paper investigates the usage of machine learning for end-to-end data rate prediction in vehicular 5G \ac{NSA} networks. To the best of our knowledge, this study represents the first empirical analysis of this network type in real world vehicular 5G \ac{NSA} scenarios.

%
%
The remainder of the paper is structured as follows. After discussing the related work in Sec.~\ref{sec:related_work}, an overview of the methodological setup for the real world data acquisition is given in Sec.~\ref{sec:methods}. Afterwards, an in-depth study of machine learning-based data rate prediction is given in Sec.~\ref{sec:results}.

\section{Related Work} \label{sec:related_work}

%
%
Although in theory, the interplay of the different mechanisms within cellular communication networks can be described deterministically, the massive system complexity typically does not allow to derive accurate analytical models of end-to-end processes \cite{Sliwa/Wietfeld/2019c}. Machine learning methods are inherently capable of closing this gap as they allow to uncover and exploit hidden dependencies between measurable indicators.
As a consequence, they have started to penetrate all areas related to wireless communications. A comprehensive summary of applications, methods, and challenges is provided by Wang et al. in \cite{Wang/etal/2020a}.

%
%
Client-based data rate prediction can be either performed \emph{actively} or \emph{passively}. 
%
%
Active prediction approaches such as \emph{LinkForecast} \cite{Yue/etal/2018a} rely on continuous time-series analysis of ongoing data transmissions that are usually exclusively injected for performing the measurements. However, vehicle-to-cloud communications is typically non-continuous and is characterized by event-controlled medium access patterns (e.g., through timer mechanisms or opportunistic methods).
%
%
As analyzed by Raida et al. in \cite{Raida/etal/2019a}, another issue is that active prediction approaches can be impacted by self-interference due to distortion of network context measurements such as \ac{RSRQ}.
%
%
Consequently, passive prediction approaches, which only rely on passively acquirable indicators, have achieved wider adoption in the scientific community. 
%
%
To the best of our knowledge, the only empirical study of client-based 5G data rate prediction has been published by Narayanan et al. in  \cite{Narayanan/etal/2020a}. However, the authors focus on the \ac{mmWave} frequency range which requires the consideration of technology-specific impact factors such as the relative angle between the antenna panels and the \acp{UE}.
%
%
Several research works have addressed passive data rate prediction within \ac{LTE} networks. Due to the continuity of the \ac{3GPP} standardization of the physical layer measurements \cite{3GPP/2020a}, these studies are still highly relevant for 5G communications. We summarize the main conclusions from literature  as follows:
%
%
\begin{itemize}
	%
	%
	\item Measurable \textbf{context indicators} such as signal strength and signal quality are highly correlated to the end-to-end data rate. However, simple prediction models using single indicators (e.g., \ac{SINR}) do not yield accurate prediction results as they only consider parts of the implied effects. In contrast to that, machine learning methods are able to exploit the joint knowledge of all locally measurable indicators \cite{Raca/etal/2020a, Nikolov/etal/2020b, Sliwa/Wietfeld/2019c}.
	
	%
	%
	\item The integration of knowledge about the \textbf{payload size} of data packets allows to implicitly consider hidden cross-layer dependencies (e.g., between transport layer mechanisms and the channel coherence time) within the predictions models \cite{Sliwa/Wietfeld/2019c}.
	
	%
	%
	\item Assuming a proper model tuning, there are only \textbf{minor differences between the classes} of machine learning models. Nevertheless, tree-based approaches such as \acf{RF} often outperform more complex methods such as (deep) \ac{ANN} which suffer from the \emph{curse of dimensionality}. In addition, they often also allow for less complex model tuning \cite{Park/etal/2019a, Herrera-Garcia/etal/2019a, Nikolov/etal/2020b, Sliwa/Wietfeld/2019c}. 
	
	%
	%
	\item A major limitation for the achievable accuracy of client-based methods is that \acp{UE} are mostly unaware of the intra-cell \textbf{traffic load} \cite{Raida/etal/2020a}. As demonstrated in \cite{Sliwa/etal/2020b}, cooperative data rate prediction, which incorporates network infrastructure knowledge, is a promising approach for improving the achievable prediction accuracy.
\end{itemize}

%
%
For the future network evolution beyond 5G \cite{Ali/etal/2020a}, there is already a consensus that \emph{pervasive intelligence} will be one of the key drivers.
%
%
In this context, the high grade of platform heterogeneity will require to adopt machine learning models with respect to the platform-specific resource constraints such as the computational power and the achievable memory, and energy efficiency \cite{Sliwa/etal/2020c}.

\section{Real World Data Acquisition Methodology} \label{sec:methods}

%
%
\begin{figure}[]  	
	\vspace{0cm}
	\centering		  
	\includegraphics[width=1.0\columnwidth]{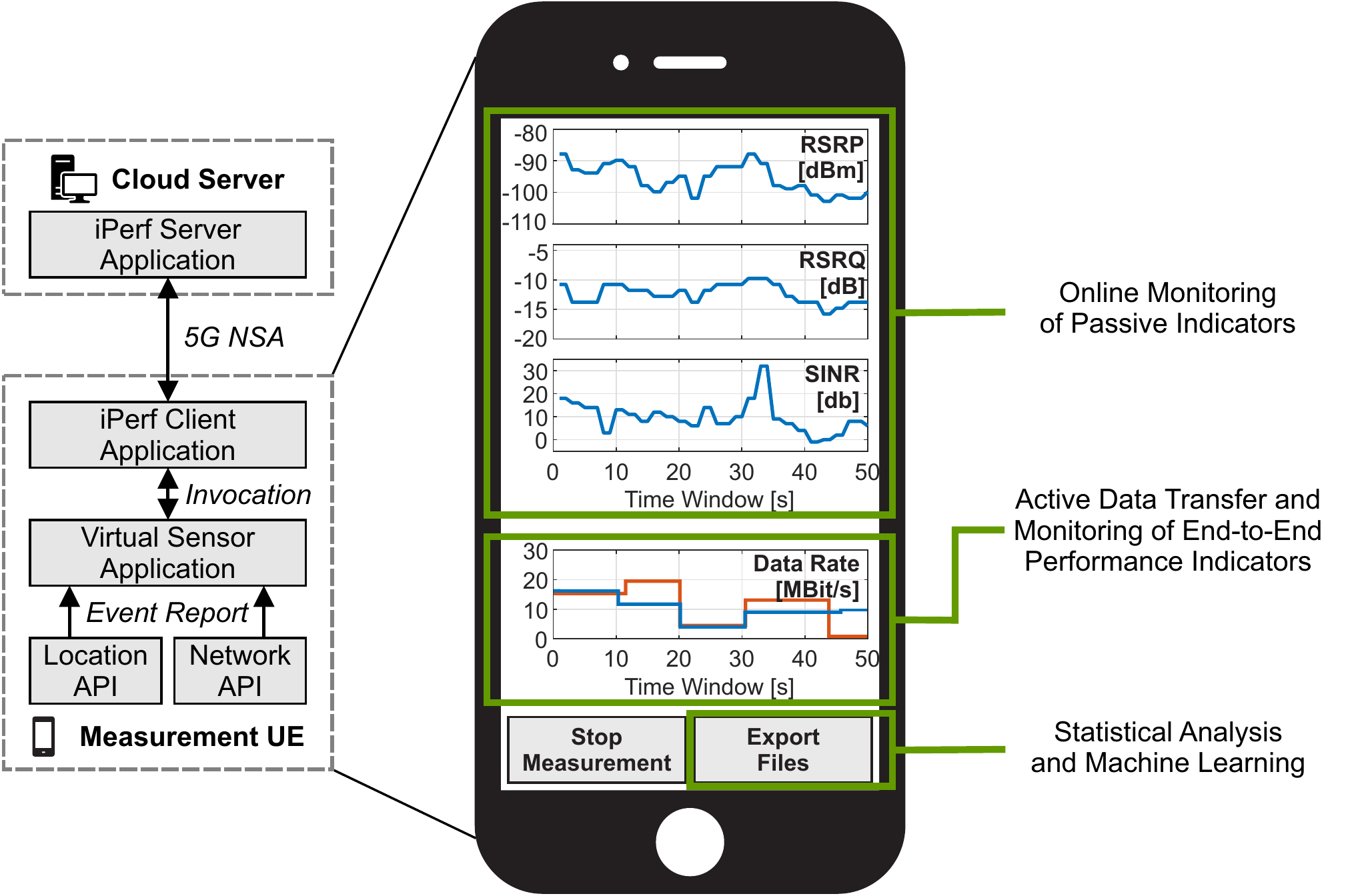}
	\caption{Illustration of the novel \texttt{Android} measurement application.}
	\label{fig:measurement_app}
	\vspace{0cm}	
\end{figure}
%
%
For performing the real world measurements, we developed a novel \texttt{Android} application (see Fig.~\ref{fig:measurement_app}) that uses the native \ac{API} to monitor \texttt{CellSignalStrengthNr} updates that report signal strength and signal quality indicators. In addition, the novel application integrates \texttt{iperf 3.9} for performing data rate measurements of active data transmissions in uplink and downlink direction.
%
%
The latter are performed with a fixed interval of \SI{10}{\second}, whereas a random payload size is selected from \SIrange{1}{10}{\mega\byte} in order to pay attention to cross-layer dependencies (e.g., between the implied transmission duration related to the payload size and the channel coherence time). Each active measurement is annotated with the corresponding context indicators (see Sec.~\ref{sec:results}).

%
%
The selection of an adequate measurement device is a non-trivial task as not all chipsets report the required context measurements to the user space. After an initial investigation, we selected the \oneplus (Qualcomm Snapdragon 865 \ac{SoC}) as a measurement device.
%
%
For completeness, it is remarked that we initially considered the usage of additional measurement devices Huawei Mate 20 X 5G, Samsung Galaxy S10 5G, and Samsung A90 5G. However, here we encountered the problem that the chipset of these devices did not support a specific \ac{NSA} frequency combination where \ac{LTE} operates at \SI{1800}{\mega\hertz} and 5G is applied at \SI{2100}{\mega\hertz}. As this frequency combination is widely utilized by the analyzed \ac{MNO} in the considered scenarios, the 5G coverage was massively reduced (from \SI{78}{\percent} to \SI{54}{\percent} in the \dortmund scenario) which rendered the usage of these devices practically unusable.

%
%
\begin{figure}[]  	
	\vspace{0cm}
	\centering		  
	\includegraphics[width=0.8\columnwidth]{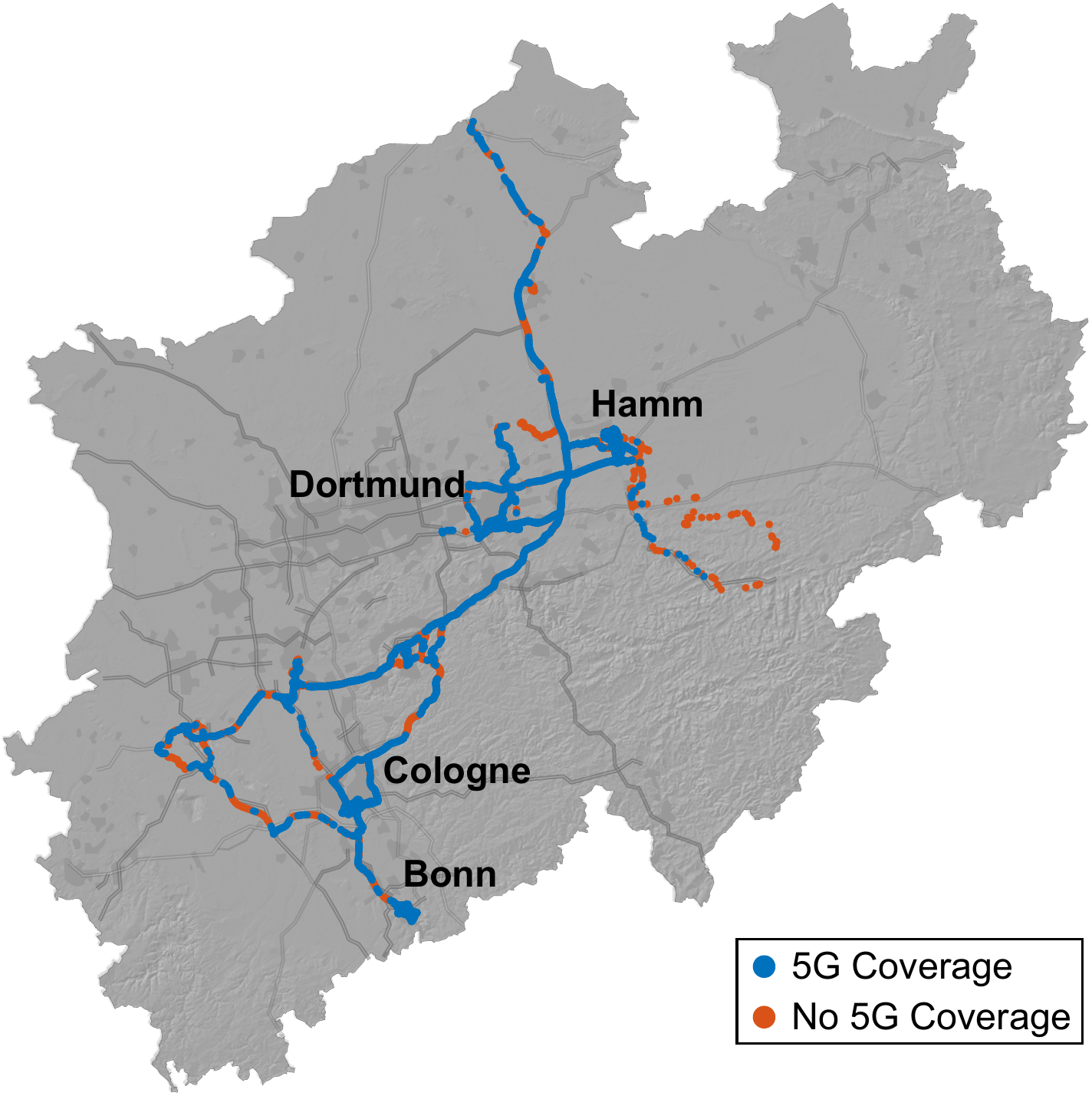}
	\caption{Overview of the measurement locations and the respective 5G coverage within the German federal state of North Rhine-Westphalia.}
	\label{fig:map_nrw}
	\vspace{0cm}	
\end{figure}
%
%
As illustrated in Fig.~\ref{fig:map_nrw}, the real world measurements cover campus, suburban, urban, and highway areas in the German cities \bonn, \cologne, \dortmund, and \hamm. The overall data set consists of 8349 active measurements and \SI{43.12}{\giga\byte} of transmitted data.
%
%
We also compared our measurements with the official statements of the \ac{MNO} concerning the 5G coverage within the considered scenarios. Without further consideration of the intended receiver sensitivity, it turned out that the \ac{MNO} optimistically claimed to achieve \SI{92.82}{\percent} 5G coverage while our measurement only achieved \SI{72.78}{\percent}. 

%
%
For strengthening the reproducibility and the reusability of the empirical results, we provide the source code of the developed application and the obtained raw measurements in an Open Source manner\footnote{The raw measurements are available at \url{https://github.com/hendrikschippers/CNI-Cell-Tracker}}.

\section{Client-Based End-to-End Data Rate Prediction for 5G NSA Networks} \label{sec:results}

%
%
Data rate prediction is a regression task. Hereby, a machine learning model $f_{\text{ML}}$ is trained offline using $N$ \emph{feature} vectors --- the context measurements --- $\mathbf{X} = \left[ \mathbf{x}_0, ..., \mathbf{x}_{N} \right]^\intercal$ with corresponding \emph{labels} --- the data rate measurements --- $\mathbf{y}$ such that $f_{\text{ML}} : \mathbf{X} \rightarrow \mathbf{y}$. Afterwards, the trained model can be utilized to make predictions $\tilde{y}$ for unlabeled context measurements $\mathbf{x}$ as $\tilde{y} = f_{\text{ML}}(\mathbf{x})$.

%
%
Each feature vector $\mathbf{x} = \left( \mathbf{x}_{\text{net}}, \mathbf{x}_{\text{mob}}, \mathbf{x}_{\text{app}}\right) $ is composed of individual context measurements from three different logical domains:
%
%
\begin{itemize}
	%
	%
	\item \textbf{Network context} $\mathbf{x}_{\text{net}}$: \ac{RSRP}, \ac{RSRQ}, \ac{SINR}, SS-RSRP, SS-RSRQ, SS-SINR, \ac{TA}, Carrier frequency
	%
	%
	\item \textbf{Mobility context} $\mathbf{x}_{\text{mob}}$: Velocity, Cell ID
	%
	%
	\item \textbf{Application context} $\mathbf{x}_{\text{app}}$: Payload size of the data packet to be transmitted
\end{itemize}
%
%
It is remarked that for \ac{RSRP}, \ac{RSRQ}, and \ac{SINR}, two variants are utilized. The regular version denotes the \ac{RS} indicators of the link to the \ac{eNB} master node whereas the \ac{SS} variants refer to the connection to the \ac{gNB} secondary node according to \cite{3GPP/2020a}.

%
%
For performing the actual machine learning-based data rate prediction, different regression models, for which the hyperparamters are determined using an initial grid search optimization, are applied. 
%
%
\begin{itemize}
	%
	%
	\item \textbf{\acf{LR}} 
	
	%
	%
	\item \textbf{\acf{ANN}} \cite{Goodfellow/etal/2016a} using the Adam optimizer and \ac{ReLU} activation function
	%
	%
	\item \textbf{\acf{RF}}	\cite{Breiman/2001a} with active pruning
	%
	%
	\item \textbf{\acf{SVM}} \cite{Smola/Schoelkopf/2004a} using the \ac{RBF} kernel
	%
	%
	\item \textbf{\acf{XGB}} \cite{Chen/Guestrin/2016a}	
\end{itemize}

%
%
\begin{table}[ht]
	\centering
	\caption{Hyperparameters of the machine learning models}
	\begin{tabular}{lp{2.5cm}cccc}
		\toprule
		
		\multirow{2}{0.5cm}{\textbf{Model}} & \multirow{2}{1.0cm}{\textbf{Hyperparameter}} & \multicolumn{2}{c}{\textbf{TCP}} & \multicolumn{2}{c}{\textbf{UDP}} \\
		&  & \textbf{UL} & \textbf{DL} & \textbf{UL} & \textbf{DL} \\
		\midrule
		
		%
		%
		\multirow{2}{0.5cm}{\textbf{LR}} 
		& Fit intercept & True & True & True & True \\
		& Normalize & True & True & False & True \\
		\midrule

		%
		%
		\multirow{4}{0.5cm}{\textbf{ANN}} 	
		& Learning rate $\eta$ & 0.2 & 0.01 & 0.01 & 0.01 \\
		& Momentum $\alpha$ & 0.001 & 0.001 & 0.3 & 0.3 \\
		& Network architecture & {[10,5]}  & {[10,10,10]} & {[10,5]} & {[10,5]} \\
		& Number of epochs & 2000 & 1000 & 1000 & 1000 \\
		\midrule
		
		%
		%
		\multirow{2}{0.5cm}{\textbf{RF}} 	
		& Max depth & 100 & 100 & 31 & 100 \\
		& Number of trees & 406 & 203 & 406 & 203 \\
		\midrule
		
		%
		%
		\multirow{2}{0.5cm}{\textbf{SVM}} 	
		& Regularization $C$ & 100 & 100 & 100 & 100 \\
		& Kernel coefficient $\gamma$ & Scale & Auto & Auto & Auto \\
		\midrule
		
		%
		%
		\multirow{4}{0.5cm}{\textbf{XGB}} 
		& Max depth & 1 & 10 & 10 & 1 \\
		& Number of trees & 1000 & 10 & 100 & 361 \\	
		& Min loss reduction $\gamma$ & 0.00 & 6.66 & 0.00 & 0.00 \\
		& Regularization $\lambda$ & 0.25 & 1.0 & 0.75 & 0.5 \\
		\bottomrule
			
	\end{tabular}
	
	\vspace{0.1cm}
	\emph{UL}: Uplink, \emph{DL}: Downlink

	\label{tab:parameters}
\end{table}

A summary of the protocol- and transmission direction-specific hyperparameter values is given in Tab.~\ref{tab:parameters}.
%
%
All machine learning evaluations are performed with the \texttt{scikit-learn} \cite{Pedregosa/etal/2011a} toolkit.

%
%
In order to evaluate the performance of the prediction models, we consider the \ac{RMSE} that is computed as
%
%
\begin{equation*}
	\text{RMSE} = \sqrt{\frac{\sum_{i=1}^{N} \left(  \tilde{y}_{i} - y_{i} \right)^2}{N}}
\end{equation*}
with $\tilde{y}$ being the prediction with corresponding ground truth measurement $y$ of the $N$ measurements.  For achieving a better understanding of the generalizability of the achieved results, we perform 10-fold cross validation.

%
%
\subsection{Comparison of Machine Learning Models}

%
%
\begin{figure}[] 
	\centering
	\subfloat[TCP]{\includegraphics[width=0.24\textwidth]{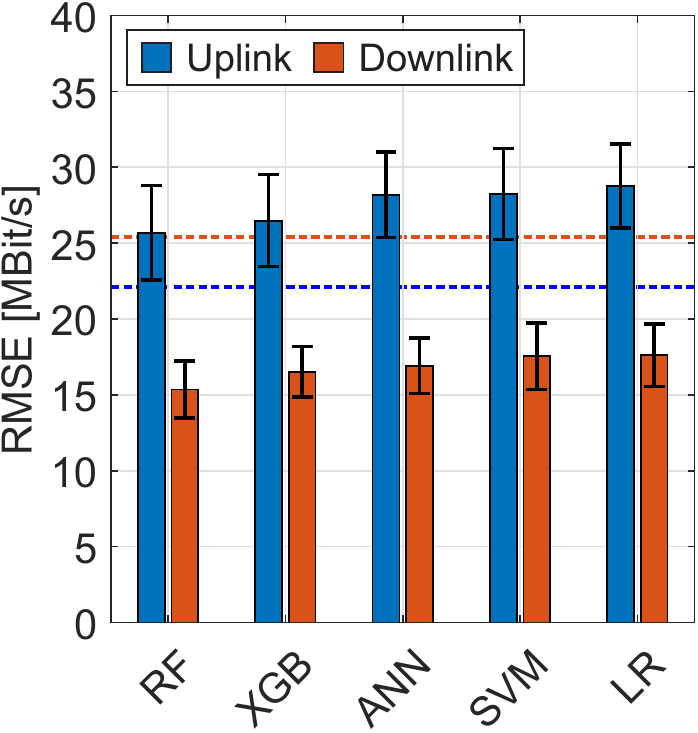}}\hfill
	\subfloat[UDP]{\includegraphics[width=0.24\textwidth]{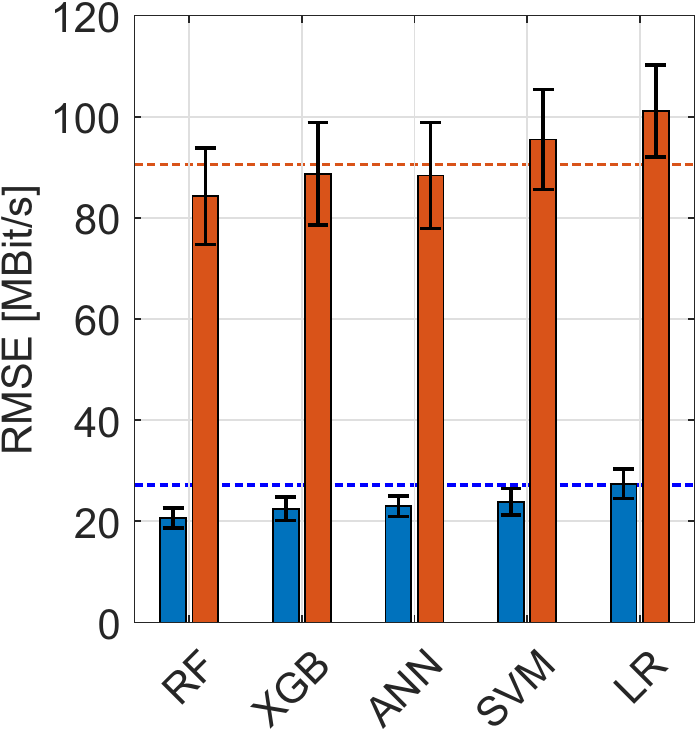}}\hfill		
	\caption{Comparison of the achieved data rate prediction accuracy for different machine learning models, transmission directions, and protocols. For reference, the dashed lines illustrate the  \SI{10}{\percent} level of the direction-specific data rate. The errorbars show the standard deviation over the different runs of the 10-fold cross validation.}
	\label{fig:cv_bars}
\end{figure}
\begin{figure}[] 
	\centering
	\subfloat[TCP (Uplink)]{\includegraphics[width=0.24\textwidth]{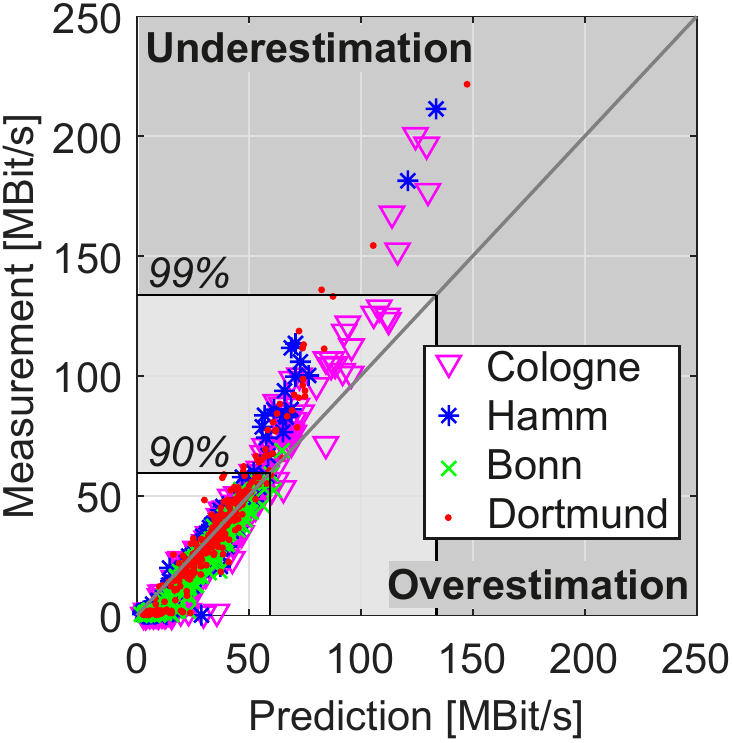}}\hfill
	\subfloat[TCP (Downlink)]{\includegraphics[width=0.24\textwidth]{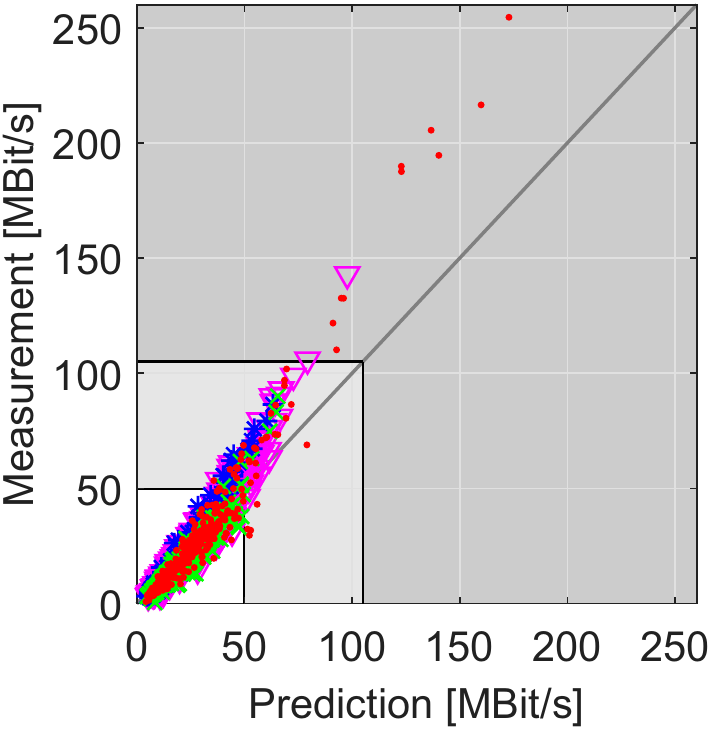}}\hfill	
	\subfloat[UDP (Uplink)]{\includegraphics[width=0.24\textwidth]{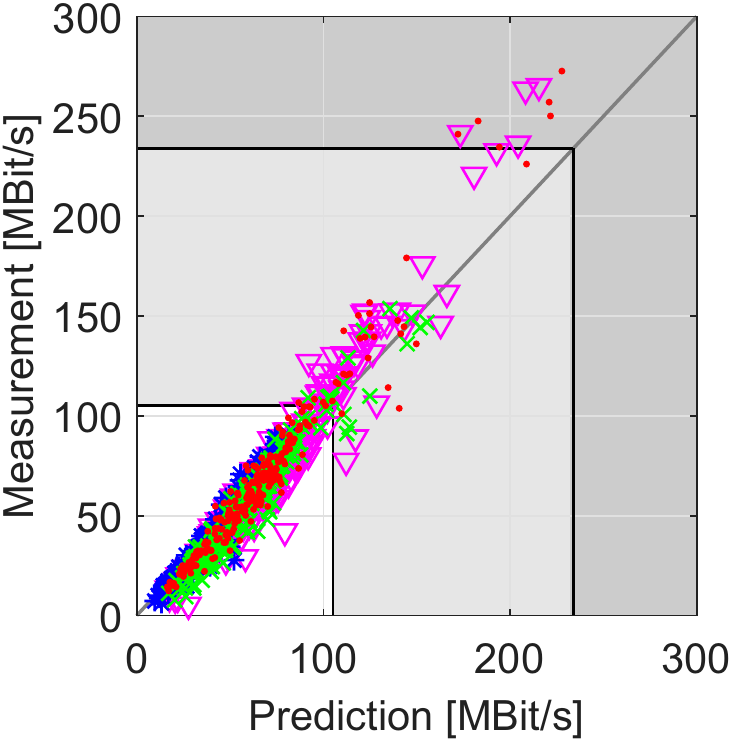}}\hfill
	\subfloat[UDP (Downlink)]{\includegraphics[width=0.24\textwidth]{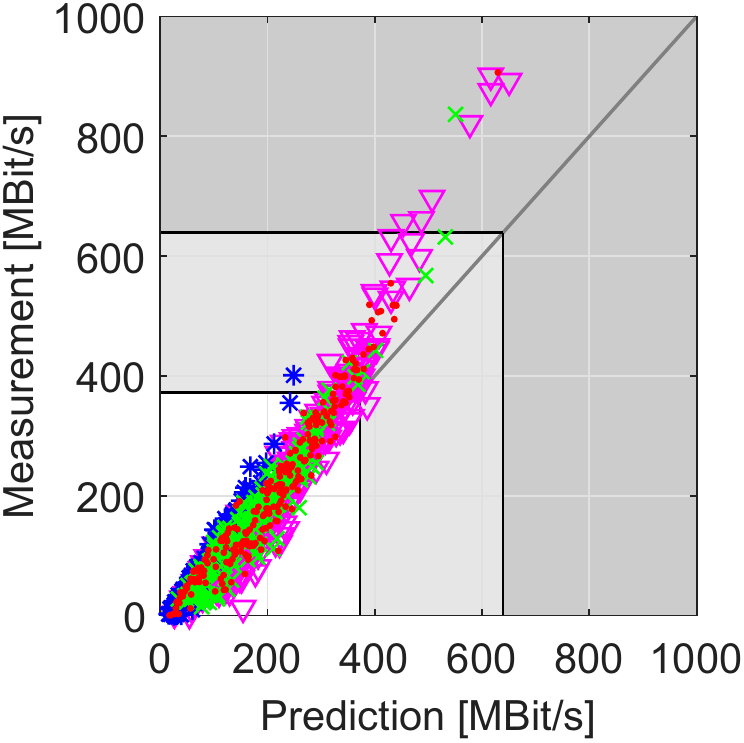}}\hfill		
	\caption{Comparison of the \ac{RF}-based data rate predictions and corresponding measurements. The diagonal lines show the behavior of hypothetically perfect prediction models. In addition, the \SI{90}{\percent} and \SI{99}{\percent} percentiles of the empirical measurements are shown.}
	\label{fig:scatterplots}
\end{figure}

%
%
At first, the performance of the regression models is compared. The results of the 10-fold cross validation are shown in Fig.~\ref{fig:cv_bars}.
%
%
As the \ac{RMSE} should be assessed relatively to the respective value range of the target variables (also see Fig.~\ref{fig:scatterplots}), the overlayed dashed lines show the \SI{10}{\percent} level of the measured data rate values as reference. In general, \udp shows larger \ac{RMSE} values than \tcp since higher peak data rates are achieved for both directions. Depending on transport protocols and transmission directions, the \ac{RF} \ac{RMSE} can be approximated as \SIrange{6}{11.5}{\percent} of the maximum data rate.
%
%
While in all variants, the lowest \ac{RMSE} is achieved by the \ac{RF} model, there are only minor differences between the (well-tuned) prediction models. It is remarkable that even simple approaches such as \ac{LR} are able to achieve a comparably high prediction accuracy. For completeness, it is remarked that minor changes of the hyperparameters of the \ac{ANN} and \ac{SVM} can lead to completely different prediction results. In contrast to that, the tree-based models allow for more intuitive and less sensitive parameter optimization. With respect to its outstanding performance, the further paragraphs focus on providing additional information for the \ac{RF} model.

%
%
%
%
The scatterplots in Fig.~\ref{fig:scatterplots} illustrate the scenario-specific distributions of the predictions and measurements for the different variants of the \ac{RF} model.
%
%
This visualization also shows that global metrics such as \ac{RMSE} are only able to provide a limited insight into the behavior of the prediction models. The 5G \ac{NSA} operation mode shows a very dynamic behavior where the peak data rates are approached infrequently. As a consequence, the prediction models show a pessimistic behavior: As an example, for the \ac{TCP} uplink model, \SI{90}{\percent} of the measurements are below \SI{27}{\percent} and \SI{99}{\percent} are below \SI{60}{\percent} of the maximum data rate.
%
%
In addition, the generalizability of the prediction model is constrained by the high grade of network heterogeneity over the different evaluation scenarios. Since the highest uplink data rates in \cologne are not observed in other scenarios, they would not be predicted if the \cologne data was removed from the training set.

%
%
\subsection{Cross-Scenario Model Generalization}

%
%
In order to analyze the generalizability of \ac{RF}-based data rate prediction, a cross-scenario performance evaluation is performed. For this purpose, the overall data set $\mathcal{D}$ is split into the $i$ scenario-specific subsets. Within each iteration, $\mathcal{D}_i$ is chosen as the test set $\mathcal{D}_{\text{train}}$ and the remaining subsets jointly form the training set $\mathcal{D}_{\text{train}}$.
%
%
\begin{figure}[] 
	\centering
	\subfloat[TCP (Uplink)]{\includegraphics[width=0.24\textwidth]{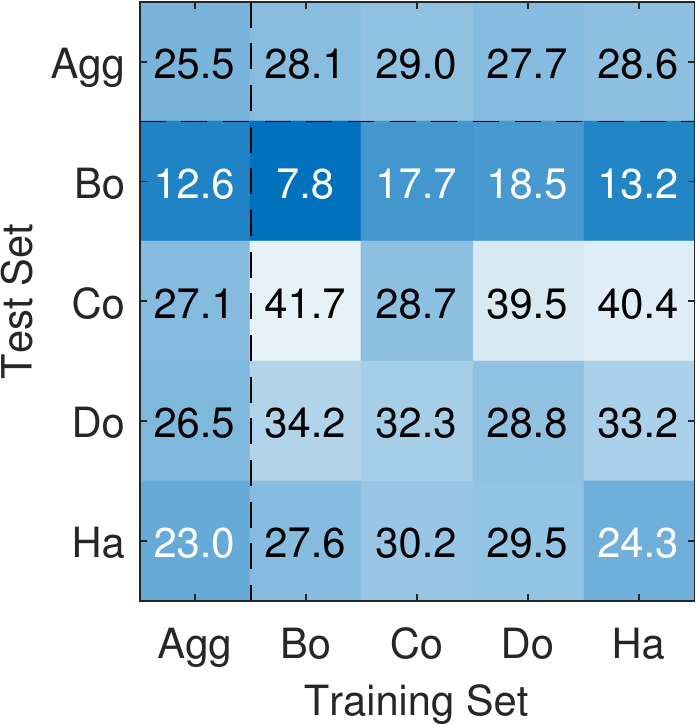}}\hfill
	\subfloat[TCP (Downlink)]{\includegraphics[width=0.24\textwidth]{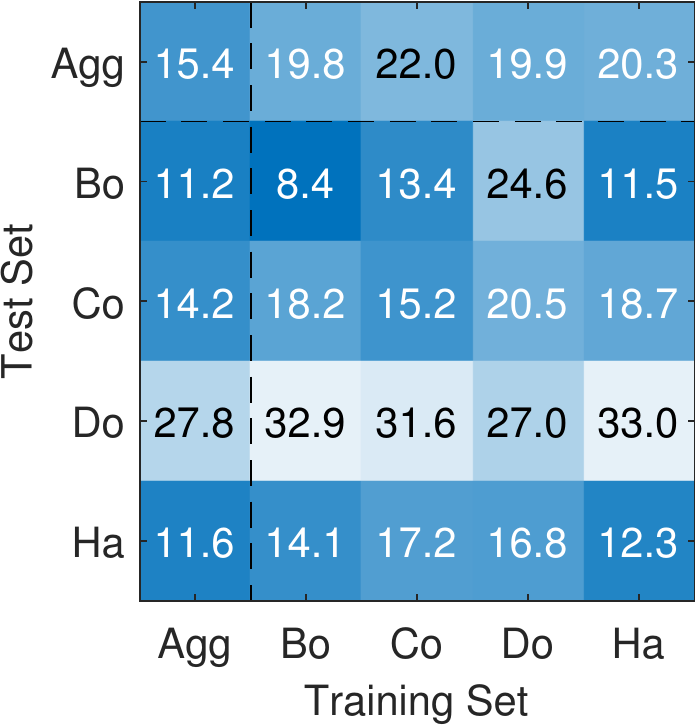}}\hfill		
	
	\subfloat[UDP (Uplink)]{\includegraphics[width=0.24\textwidth]{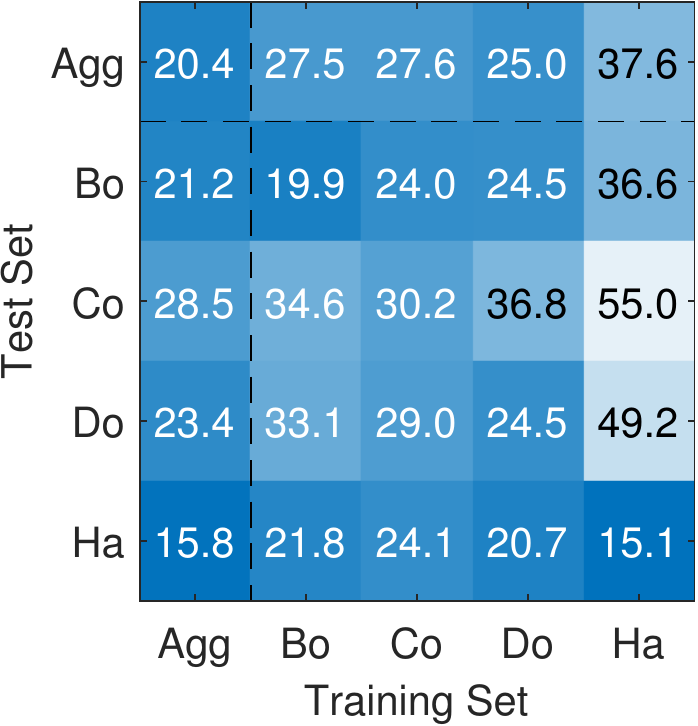}}\hfill
	\subfloat[UDP (Downlink)]{\includegraphics[width=0.24\textwidth]{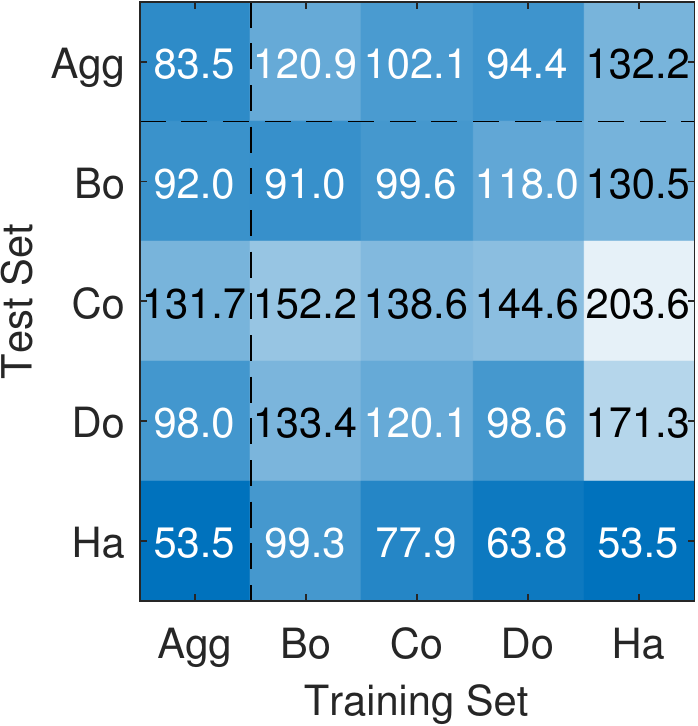}}\hfill	
	
	\caption{\ac{RMSE} cross-scenario performance evaluation of the \ac{RF} models. \emph{Agg}: Aggregated data of all scenarios, \emph{Bo}: Bonn, \emph{Co}: Cologne, {Do}: Dortmund, \emph{Ha}: Hamm.}
	\label{fig:cross_scnenario}
\end{figure}
%
%
The resulting \ac{RMSE} matrices for the transport protocols and transmission directions are shown in Fig.~\ref{fig:cross_scnenario}.
%
%
For each variant, the first column shows the performance of the overall prediction model in each of the individual scenarios. In this case, the aggregated training set is composed of \SI{90}{\percent} of the data per scenario and the remaining \SI{10}{\percent} forms the test set.

%
%
By comparing the values of the first column $\mathbf{R}_1$ of the \ac{RMSE} matrix $\mathbf{R}$ with the main diagonal $\mathbf{R}_{\text{diag}} = \diag^{-1}(\mathbf{R})$, it can be determined if the model leverages the additional data of the other scenarios for achieving better predictions ($\mathbf{R}_1 < \mathbf{R}_{\text{diag}}$) or if it rather benefits from applying a more local perspective ($\mathbf{R}_{\text{diag}} < \mathbf{R}_1$).
%
%
For all variants, the results are ambiguous and no clear trend can be identified. It is remarked that these findings are partially opposed to our previous analysis of \ac{LTE} communications in \cite{Sliwa/Wietfeld/2019c} where we --- in compliance with general data science principles \cite{Domingos/2012a} --- concluded that the machine learning models benefit more from additional data than from applying a more local perspective. However, the 5G \ac{NSA} results illustrate that in some cases, heterogeneous and non-converged networks --- the 5G market penetration is still in the early adoption phase ---  require a deeper consideration of the local phenomena.

%
%
Analogously, the first row $\mathbf{R}^\intercal_{1}$ shows the capability of each scenario-specific model to induce the behavior of the overall data set. While none of scenario-specific models is able to approach the performance of the cross-validated aggregated data set, there are also significant differences between the scenarios. Here, the highest average generalization is achieved by the \dortmund scenario. These findings illustrate that machine learning models for end-to-end data rate prediction require a high grade of versatility of the features and labels. As an example, a model that is only trained on the \bonn scenario is unable to predict the higher data rates that occur in the \cologne scenario.

%
%
\subsection{Relative Feature Importance}

%
%
\begin{figure}[] 
	\centering
	\subfloat[TCP]{\includegraphics[width=0.24\textwidth]{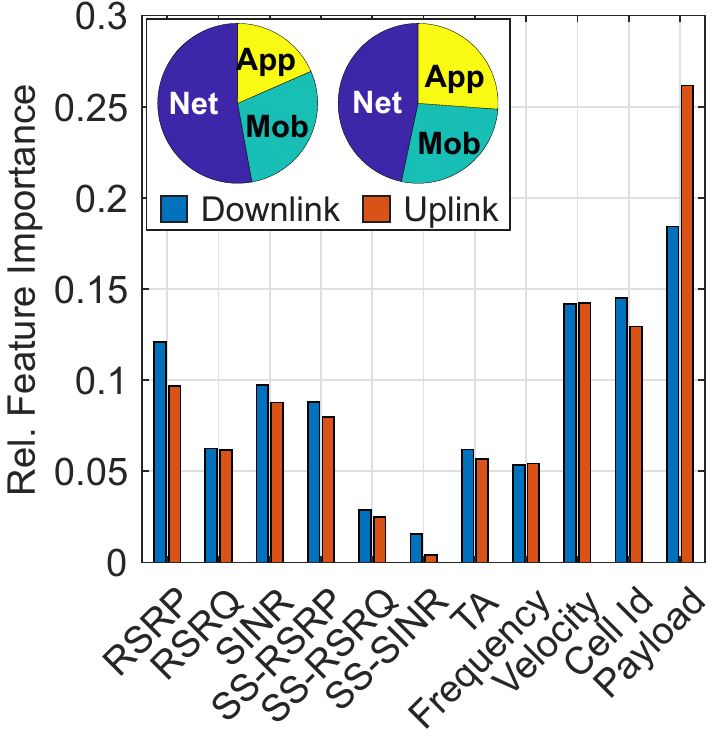}}\hfill
	\subfloat[UDP]{\includegraphics[width=0.24\textwidth]{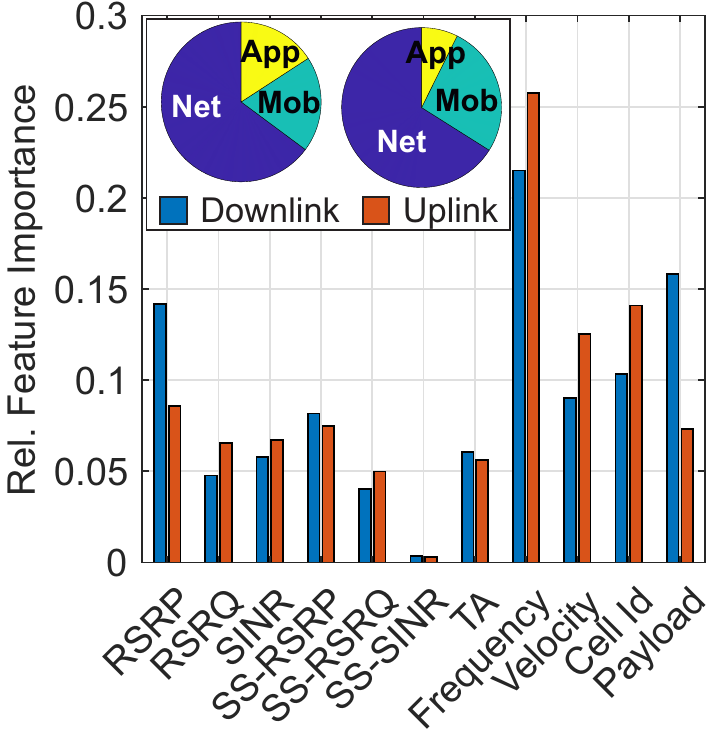}}\hfill		
	\caption{Relative feature importance of the \ac{RF} model based on \ac{MDI} analysis. The overlayed pie chart illustrates the overall feature importance per context domain.}
	\label{fig:feature_importance}
\end{figure}
In order to gain insight into the interplay of the different features of the \ac{RF} model, the \ac{MDI}-based relative feature importance \cite{Louppe/etal/2013a} is analyzed in Fig.~\ref{fig:feature_importance}.
It remarked that the interpretation of these results is non-trivial: The \ac{MDI} only represents the internal description of the measurements by the \ac{RF} model which does not necessarily equal the underlying real world impact factors.
%
%
The results show a high similarity between the two transmission directions, which indicates that network quality is often symmetric and that passive downlink indicators provide meaningful information for estimating the uplink behavior.
%
%
Although for both protocols, the network context $\mathbf{x}_{\text{net}}$ is the most significant context domain, the payload size of the data packet is the most important individual feature for \ac{TCP}. This observation is in compliance with earlier \ac{LTE} measurements of \cite{Sliwa/Wietfeld/2019c} and can be explained by the slow start mechanism and potential retransmissions.
%
%
In contrast to that, the network context features have an even increased significance for \ac{UDP} with the carrier frequency being the most relevant individual feature. The latter can be regarded as a weighting factor that scales the impact of the other network context features.
%
%
In all variants, the importance of the SS-SINR is very low as this indicator is rarely contained in the measurement reports of the \ac{UE}. However, as the respective calculation and reporting mechanisms are not standardized by \ac{3GPP} and depends on the modem manufacturer, this observation might be specific for the considered measurement device.

\section{Conclusion}

%
%
In this paper, we presented an empirical analysis of client-based data rate prediction in vehicular 5G \ac{NSA} scenarios that covers \ac{TCP} and \ac{UDP} in uplink and downlink direction. 
%
%
Although 5G \ac{NSA} already delivers \ac{eMBB} peak data rates, a large amount of data rate measurements occur in the lower regions. This dynamic behavior renders 5G data rate prediction a challenging task and illustrates that the \ac{NSA} operation mode is not yet able to provide the \ac{QoS} guarantees for highly resource demanding services such as teleoperation.
%
%
However, our analysis also shows that the general principles of machine learning-based data rate prediction that utilizes context measurements as input features are still applicable. In all variants, the \ac{RMSE} can be approximated as \SIrange{6}{11.5}{\percent} of the \ac{MNO}- and protocol-specific maximum data rate.
%
%
It is remarked that the 5G adoption rate is in the early market stage and the chasms towards the mainstream market is yet to be crossed. Thus, we will monitor the further developments towards a more converged network state, especially in the light of spectral resource competition implied by increasing number of active users. Moreover, we will extend our data set with measurements for other \acp{MNO} and evaluation scenarios. 
\ifdoubleblind

\else

	\section*{Acknowledgment}
	
	\footnotesize
	%
	%
	This work has been supported by the German Research Foundation (DFG) within the \emph{Collaborative Research Center SFB 876} ``Providing Information by Resource-Constrained Analysis'', projects A4 and B4,
	%
	%
	as well as by the Ministry of Economic Affairs, Innovation, Digitalization and Energy of the state of North Rhine--Westphalia (MWIDE NRW) in the course of the \emph{Competence Center 5G.NRW} under grant number 005--01903--0047
	%
	%
	and has received funding by the Federal Ministry of Transport and Digital Infrastructure (BMVI) in the context of the project \emph{Virtual integration of decentralized charging infrastructure in cab stands} under the funding reference 16DKVM006B.

\fi

\balance

\ifacm
\else
	\bibliographystyle{IEEEtran}
	\bibliography{Bibliography}
\fi

\end{document}